\documentclass[aps,prd,twocolumn,superscriptaddress,amssymb,eqsecnum,showpacs,showkeyes,secnumarabic,graphics,floatfix,nofootinbib,tightenlines,longbibliography]{revtex4-1}
\usepackage{relsize}
\usepackage{blindtext}
\usepackage{enumitem}
\usepackage{xcolor}
\usepackage{graphicx} 
\usepackage{epsfig}
\usepackage{bm} 
\usepackage{cancel}
\usepackage{dcolumn}
\usepackage{amsmath} 
\usepackage{hhline} 
\usepackage[utf8]{inputenc}
\usepackage[breaklinks=true,colorlinks=true,
linkcolor=blue,urlcolor=blue,citecolor=cyan,
bookmarks=true,bookmarksopenlevel=2]{hyperref}

\def \beq  {\begin{equation}}
\def \eeq  {\end{equation}}
\def \ber  {\begin{eqnarray}}
\def \eer  {\end{eqnarray}}

\begin{document}
\newcommand{\newc}{\newcommand}

\newc{\be}{\begin{equation}}
\newc{\ee}{\end{equation}}
\newc{\ba}{\begin{eqnarray}}
\newc{\ea}{\end{eqnarray}}
\newc{\bea}{\begin{eqnarray*}}
\newc{\eea}{\end{eqnarray*}}
\newc{\D}{\partial}
\newc{\ie}{{\it i.e.} }
\newc{\eg}{{\it e.g.} }
\newc{\etc}{{\it etc.} } 
\newc{\etal}{{\it et al.}}
\newc{\lcdm}{$\Lambda$CDM }
\newcommand{\nn}{\nonumber}
\newc{\ra}{\Rightarrow}

\title{Scalar tachyonic instabilities in gravitational backgrounds: Existence and growth rate} 
\author{L. Perivolaropoulos}\email{leandros@uoi.gr} 
\author{F. Skara}\email{f.skara@uoi.gr}
\affiliation{Department of Physics, University of Ioannina, 45110 Ioannina, Greece}

\date {\today} 

\begin{abstract}
It is well known that the Klein Gordon (KG) equation $\Box \Phi + m^2\Phi=0$  has tachyonic unstable modes on large scales ($k^2<\vert m \vert^2$) for $m^2<m_{cr}^2=0$ in a flat Minkowski spacetime with maximum  growth rate $\Omega_{F}(m)= \vert m \vert$ achieved at $k=0$. We investigate these instabilities in a Reissner-Nordström-deSitter (RN-dS) background spacetime with mass $M$, charge $Q$, cosmological constant $\Lambda>0$ and multiple horizons. By solving the KG equation in the range between the event and cosmological horizons, using tortoise coordinates $r_*$, we identify the bound states of the emerging Schrodinger-like Regge-Wheeler equation corresponding to instabilities. We find that the critical value  $m_{cr}$ such that for $m^2<m_{cr}^2$ bound states and instabilities appear, remains equal to the flat space value $m_{cr}=0$ for all values of background metric parameters despite the locally negative nature of the Regge-Wheeler potential for $m=0$. However, the growth rate $\Omega$ of tachyonic instabilities for $m^2<0$ gets significantly reduced compared to the flat case for all parameter values of the background metric ($\Omega(Q/M,M^2 \Lambda, mM)< \vert m \vert$). This increased lifetime of tachyonic instabilities is maximal in the case of a near extreme Schwarzschild-deSitter (SdS) black hole where $Q=0$ and the cosmological horizon is nearly equal to the event horizon ($\xi \equiv 9M^2 \Lambda \simeq 1$).  The physical reason for this delay of instability growth appears to be the existence of a cosmological horizon that tends to narrow the negative range of the Regge-Wheeler potential in tortoise coordinates. 
\end{abstract}
\maketitle
  
\section{Introduction}  
\label{sec:Introduction}  

Scalar fields are used to describe a wide range of degrees of freedom in a diverse set of physical systems in particle physics (e.g. the Higgs field and other symmetry breaking scalar fields\cite{Susskind:1978ms}), cosmology (e.g. the inflaton \cite{Linde:1983gd} and the quintessence field\cite{Zlatev:1998tr}), gravitational theories (e.g. scalar field hair on black holes\cite{Herdeiro:2015waa} or modified gravity scalar degrees of freedom like $f(R)$ theories \cite{Sotiriou:2008rp,Carroll:2003wy,DeFelice:2010aj,Capozziello:2005ku,Nojiri:2006gh,Nojiri:2006ri,Nojiri:2007cq,Hu:2007nk,Fay:2007uy,Capozziello:2007ec,Nojiri:2010wj,Basilakos:2013nfa} or scalar tensor theories\cite{Boisseau:2000pr}), condensed matter (e.g. the Bose-Eistein scalar field condensate\cite{Anderson:1995gf}) etc.

The dynamical evolution of a scalar field in a classical system is determined by three main factors
\begin{itemize}
\item The form of its Lagrangian density and especially the scalar field potential $V(\phi)$ which may be e.g. of the form $V(\phi)=m^2 \phi^2$ for a simple massive scalar field or of a symmetry breaking form $V(\phi)=\frac{\lambda}{4}(\phi^2-\eta^2)^2$ where $\eta$ is the scale of symmetry breaking.
\item The form of the background spacetime which may be for example flat Minkowski, cosmological Friedmann-Robertson-Walker (FRW), Schwarzschild etc.
\item The boundary/initial conditions used for the solution of the resulting dynamical scalar field equation emerging from the above two factors.
\end{itemize}
The simplest Lagrangian density describing the evolution of a scalar field is that corresponding to a free massive scalar which is of the form
\ba
{\cal L}&=&\frac{1}{2}\partial_\mu\Phi \partial^\mu \Phi - m^2 \Phi^2
\label{scfieldlang}
\ea
\noindent  leading to the Klein-Gordon equation \cite{Landau:1982dva}
\be
\Box \Phi + m^2\Phi=0
\label{kgeq}
\ee
In flat Minkowski space this equation may be written as 
\be
\ddot{\Phi}-\nabla^2 \Phi=-m^2\Phi
\label{nlptwoeq1flat}
\ee
Its solutions are propagating waves of the form
\be
\Phi(\vec r,t)=A(\vec k) e^{i(\omega t -\vec k \vec r)} + B(\vec k) e^{-i(\omega t -\vec k \vec r)}
\label{phisol}
\ee
with dispersion relation
\be
\omega^2=k^2+m^2
\label{dirprel1}
\ee
For $m^2>0$ we have well behaved propagating waves. However, for $m^2<0$ we have
\be
\omega=\pm\sqrt{k^2-\vert m\vert^2}
\label{nlptwoeq3flat}
\ee
and exponentially growing tachyonic instabilities develop on large scales ($k<\vert m \vert$) where $Im(\omega)\neq 0$ \cite{Felder:2000hj}. In the context of a spontaneous symmetry breaking potential, these instabilities usually imply the presence of a broken symmetry and the transition of the scalar field to a new stable (or metastable) vacuum. However, in the context of a potential that is unbounded from below they may also imply that the theory is unphysical and should be ruled out. This argument has lead to disfavor of a wide range of theories which involve scalar fields with negative $m^2$ including a wide range of massive Brans-Dicke (BD) theories and $f(R)$ theories  where such tachyonic instabilities are also known as Dolgov-Kawasaki-Faraoni (DKF) instabilities \cite{Dolgov:2003px,Faraoni:2006sy}(see also \cite{Soussa:2003re,Kobayashi:2008tq,Nojiri:2003ft,Nojiri:2003ni,Seifert:2007fr,Sawicki:2007tf}).
For example a massive BD scalar field has an action of the form\footnote{The BD parameter $\omega$ should not be confused with angular frequency $\omega$ used above.}\cite{PhysRev.124.925,Alsing:2011er,Perivolaropoulos:2009ak,Torres:2002pe,Sen:2000vj}
\begin{widetext}
\be
S=\frac{1}{16\pi G}\int d^4 x\sqrt{-g}\left[\Phi R - \frac{\omega}{\Phi}g^{\mu \nu}\partial_{\mu}\Phi \partial_{\nu}\Phi
-m^2(\Phi-\Phi_0)^2\right]
\label{actionst}
\ee
\end{widetext}
In this theory (using finite boundary conditions at infinity)  a small point mass $M$ located at the origin creates a scalar field and metric configurations of the form
\be 
\Phi=\Phi_0+\varphi
\label{weakgravfield}
\ee
\be
g_{\mu\nu}=\eta_{\mu\nu}+h_{\mu\nu}
\label{weakgravmertric}
\ee
where
\be
\varphi=\frac{2GM}{(2\omega+3)r}e^{-\bar{m}(\omega)r}
\label{weakgravphi}
\ee
\be
h_{00}=\frac{2GM}{\Phi_0 r}\left(1+\frac{1}{2\omega +3} e^{-\bar{m}(\omega)r}\right)
\label{weakgravh00}
\ee
\be
h_{ij}=\frac{2GM}{\Phi_0 r}\delta_{ij}\left(1-\frac{1}{2\omega +3} e^{-\bar{m}(\omega)r}\right)
\label{weakgravh00}
\ee
with $\bar{m} (\omega)=\sqrt{\frac{2\Phi_0 m^2}{2\omega +3}}$ ($\Phi_0$ is dimensionless) \cite{Perivolaropoulos:2009ak}. 

This $h_{00}$ metric perturbation corresponds to an effective Newton's constant that has a Yukawa correction of the form
\be
G_{eff}=\frac{G}{\Phi_0}\left(1+\frac{1}{2\omega +3} e^{-\bar{m}(\omega)r}\right)
\label{geff}
\ee
This Yukawa correction is decaying exponentially for $m^2>0$ and is observationally/experimentally viable either for large values of $\omega>40000$ \cite{Will:2014kxa} (so that the amplitude of the Newtonian correction is small) or for large values of the scalar field mass $m$ (so that the Newtonian correction decays fast) \cite{Perivolaropoulos:2009ak}. 

For $m^2<0$ it is easy to show that the corresponding $G_{eff}$ is spatially oscillating with wavelength $\lambda \simeq \frac{2\pi}{\bar {m}} $
\be
G_{eff}=\frac{G}{\Phi_0}\left(1+\frac{1}{2\omega+3}cos\left(\bar m(\omega) r +\theta\right)\right)
\label{geffoscil}
\ee
where $\theta$ is an arbitrary constant. For spatial oscillations of $G_{eff}$ with wavelength less that sub-mm scales ($m\gtrsim 10^{-3} eV$ ($\lambda \lesssim  1mm$) \cite{Perivolaropoulos:2016ucs,Kapner:2006si}) these spatial oscillations of $G_{eff}$ would have hardly any observational/experimental effects with current experiments/observations despite of the fact that there is no Newtonian limit as $m^2\rightarrow 0^-$ \cite{Perivolaropoulos:2016ucs,Olmo:2005hc}. This is due to the local spatial cancellation of the spatially oscillating force correction. However, the main problem with $m^2<0$ are tachyonic instabilities \cite{Capozziello:2009nq,Faraoni:2008ke,Nojiri:2007jr,Cognola:2007zu}. 

It is easy to show that perturbations of the BD scalar Eq. (\ref{weakgravphi})  obey in flat space a KG equation of the form
\be
\ddot{\delta\varphi}-\nabla^2 \delta\varphi+m^2\delta\varphi=0
\label{klgormass}
\ee
which for $m^2<0$ implies the presence of exponentially growing with time tachyonic instabilities for large scales \cite{Perivolaropoulos:2016ucs}. Thus, this theory with $m^2<0$ is only viable if the unstable scales are pushed beyond the cosmological horizon $\sim H_0^{-1}$ which corresponds to scalar field mass $\vert m \vert <10^{-33}eV$ similar to a quintessence scalar field mass. Such spatially oscillating modes have a cosmological horizon scale wavelength and have no  observable effects on small scale gravity experiments.

In the case of $f(R)$ theories which may be shown to be equivalent to BD theories with no kinetic term ($\omega=0$) \cite{Chiba:2003ir,Teyssandier:1983zz,Wands:1993uu,Faraoni:2006hx,Capozziello:2010wt} a similar instability occurs. For example the $f(R)$ theory  of the form (Starobinsky model \cite{1980PhLB...91...99S})
\be
f(R)=R+\frac{1}{6m^2}R^2
\label{frmodel}
\ee
\noindent is easily shown to be equivalent to the BD theory  with action \cite{Berry:2011pb,Capozziello:2009vr,Capozziello:2007ms,Chiba:2006jp,Olmo:2006eh,Faraoni:2006hx,Perivolaropoulos:2016ucs}
\be
\begin{split}
S_{BD}&=\frac{1}{16\pi G}\int d^4 x\sqrt{-g}\left[\Phi R - \frac{3}{2}m^2(\Phi-1)^2\right]\\
&+S_{matter}
\end{split}
\label{actionbd}
\ee
and therefore has the same tachyonic instabilities as the above mentioned massive BD theory (DKF instability).

The parameter value $\vert m\vert \simeq 10^{-3}eV$ with $m^2<0$ leads to an oscillating Newton's constant with wavelength about $1mm$. In this case the lifetime of the unstable tachyonic modes in Minkowski spacetime would be about $10^{-11} sec$. Thus, even though the mass range $\vert m\vert > 10^{-3}eV$ with $m^2<0$  leads to oscillating modifications of Newton's constant that are consistent with observations/experiments, in the context of $f(R)$ and BD theories and in a flat space background, this mass range is ruled out due to the predicted tachyonic instabilities. This inconsistency is undesirable in view of recent studies \cite{Perivolaropoulos:2016ucs,Antoniou:2017mhs,Perivolaropoulos:2019vkb} \footnote{For viable theoretical models with spatially oscillating $G_{eff}$ see \cite{Conroy:2017nkc,Conroy:2014eja,Edholm:2016hbt}.} that pointed out the existence of oscillating force signals in short range gravity experiments. It is therefore interesting to investigate if there are physical conditions that can eliminate these tachyonic instabilities or at least drastically change their lifetime.

A crucial assumption used in the derivation of the above tachyonic instability is the existence of a Minkowski background. The following questions therefore emerge:
\begin{itemize}
    \item Do scalar tachyonic instabilities for $m^2<0$ persist in the presence of a non-flat background?
    \item How do the instability lifetime and growth rate change in a curved background?
    \item What are the parameter values of a background metric required to significantly increase the instability lifetime compared to its value in a Minkowski spacetime?
\end{itemize}

The main goal of the present analysis is to address these questions. In particular we solve the KG equation in a Reissner-Nordström-deSitter (RN-dS) background metric \cite{Lake:1979zzb,Laue:1977zz} with charge $Q$, mass $M$ and cosmological constant $\Lambda$, in the region between the event horizon and the cosmological horizon with boundary conditions corresponding to a finite scalar field $\Phi$ with exponential tachyonic instabilities. Using tortoise coordinates that shift these horizons to $\pm \infty$, the KG equation is reduced to a Schrodinger-like Regge-Wheeler equation whose bound states correspond to instability modes. We find the critical value of $m^2$ ($m_{cr}^2$) such that for $m^2<m_{cr}^2$ bound states (instability modes) exist. For the tachyonic unstable modes ($m^2<m_{cr}^2$) we also find the growth rate of the instabilities (ground state eigenvalues of Regge-Wheeler equation) and compare with the corresponding growth rate in a flat Minkowski background. We also consider special cases of the RN-dS metric including the Schwarzschild metric \cite{Schwarzschild:1916uq},  the deSitter (dS) metric \cite{deSitter:1916zza,deSitter:1916zz,deSitter:1917zz,Hawking:1973uf,Bousso:2002fq}, the Schwarzschild-deSitter (SdS) metric \cite{1934rtc..book.....T,Stephani:2003tm} and the  Reissner-Nordström (RN) metric \cite{1916AnP...355..106R,1918KNAB...20.1238N,Griffiths:2009dfa}.

In the present analysis we focus on the existence of tachyonic exponentially growing solutions and do not consider propagating waves on the boundary horizons which would lead to calculation of Quasinormal Modes\footnote{A semi-analytical method for calculations of QNMs based on the Wentzel-Kramers-Brillouin (WKB) approximation \cite{Schutz:1985km,Iyer:1986np}. This method was used in a wide range of spacetimes and in a lot of studies (see e.g. \cite{Konoplya:2003ii,Konoplya:2001ji,Konoplya:2003dd,Cho:2003qe,Yang:2012he,Matyjasek:2017psv,Hatsuda:2019eoj,Devi:2020uac,Churilova:2020aca,Churilova:2020mif,Lagos:2020oek}).} (QNMs) \cite{Vishveshwara:1970cc,Edelstein:1970sk} (see Refs. \cite{Kokkotas:1999bd,Nollert:1999ji,Berti:2009kk,Ferrari:2007dd,Destounis:2019zgi} for reviews on QNMs of black holes). Such investigation of QNMs has been performed in previous studies in Schwarzschild black hole \cite{Hod:1998vk,Xue:2003vs,Cardoso:2003vt}, in SdS background for $m=0$   \cite{Zhidenko:2003wq,Abdalla:2003db,Choudhury:2003wd,MaassenvandenBrink:2003yq}, for $m^2>0$ in RN-dS background \cite{Hod:2013eea,Zhu:2014sya,Hod:2018fet,Konoplya:2014lha} and in Kerr-deSitter background \cite{Detweiler:1980uk,Konoplya:2006br,Dolan:2007mj,Cardoso:2005vk,Witek:2012tr,Okawa:2014nda} where a different type of instability was observed in the context of scalar field wave scattering. This instability is connected with the phenomenon of superradiance \cite{Bekenstein:1973mi,Damour:1976kh,Konoplya:2011qq,Cardoso:2013krh,Shlapentokh-Rothman:2013ysa,Herdeiro:2013pia,Zhang:2014kna,Brito:2015oca,Sanchis-Gual:2015lje,Moschidis:2016wew} in which a reflected wave has larger amplitude than the corresponding incident wave. Superradiant instabilities occur in rotating and in charged black holes embedded in a deSitter space and are based on the extraction of mass and/or rotational or electromagnetic energy from the black hole. This energy is then carried away from the black hole during a scattering process through the propagation of a reflected scalar field wave with amplitude increased compared to the incident scalar field wave.  Superradiance would lead to a decrease in black hole energy and increase of the energy of the scalar field causing further enhancement of the instability. Thus, the endpoints of such instability could be the evacuation of matter from the black hole  and/or the formation of a novel scalar field configuration around the black hole leading to a phenomenon called 'scalarization' and violation of the no-hair theorem, which states that black holes are fully characterized by their mass, charge and angular momentum.  A crucial property of spacetimes with superradiant instabilities is the combination of an event horizon with a cosmological deSitter horizon in four or higher dimensions \cite{Zhidenko:2009zx,Destounis:2019hca}. In this context one of the goals of the present analysis is the identification of the role of this combination of horizons on tachyonic instabilities and the discussion of their possible connection with superradiant instabilities which involve boundary conditions of propagating wave modes. 

The structure of this paper is the following: In the next section \ref{kgsdsspacetime} we use spherical tortoise coordinates $r_*$ in the context of an instability ansatz, to transform the KG equation to a Schrodinger-like Regge-Wheeler equation for the radial function $u_l(r_*)$ with potential that depends on the angular scale $l$, the dimesionless parameters $\xi\equiv 9 M^2 \Lambda $ and $q\equiv Q/M$ defined above as well as the scalar field mass $m^2$. The existence of unstable modes that are finite at the two horizons, is equivalent with the existence of bound states of this Regge-Wheeler equation. In section \ref{numsol},  we solve the Regge-Wheeler equation numerically and identify the range $m^2(q,\xi)$ for which bound states (unstable modes) exist. In the parameter range that remains unstable ($m^2<m_{cr}^2(q,\xi)$) we find the growth rate $\Omega$ of the instabilities. In section \ref{singlehorizon} we discuss the scalar tachyonic instabilities in the  limiting cases of pure deSitter and pure Schwarzschild backgrounds. Finally, in section \ref{conclusion} we conclude and discuss the physical implications of our results. We also discuss possible extensions of this analysis.

In what follows we use Planck units ($G=c=\hbar=1$) and a metric signature $(+---)$.

\section{KG equation in SdS/RN-dS spacetimes}
\label{kgsdsspacetime}

\subsection{Schwarzschild-deSitter background}
Consider a SdS background spacetime defined by the metric \cite{1934rtc..book.....T}

\be
ds^2=f(r)dt^2-\frac{1}{f(r)}dr^2-r^2(d\theta^2+\sin^2\theta d\phi^2)
\label{nlptwoeq1}
\ee
where
\be
f(r)=1-\frac{2M}{r}-\frac{\Lambda}{3}r^2
\label{nlptwoeq2}
\ee
In such a background the KG equation (\ref{kgeq}) takes the form
\be 
\frac{1}{f(r)}\frac{\partial^2\Phi}{\partial t^2}-\frac{\partial}{\partial r}f(r)\frac{\partial\Phi}{\partial r}-\frac{2f(r)}{r}\frac{\partial\Phi}{\partial r}-\frac{\Delta_{\theta\phi}\Phi}{r^2}+m^2\Phi=0
\label{nlptwoeq3}
\ee
with 
\be
\Delta_{\theta\phi}=\frac{1}{\sin\theta}\frac{\partial}{\partial\theta}\sin\theta\frac{\partial}{\partial\theta}+\frac{1}{\sin^2\theta}\frac{\partial^2}{\partial\phi^2}
\label{nlptwoeq4}
\ee
Using now the ansatz
\be 
\Phi(t,r,\theta,\phi)=\sum_{lm}\frac{\Psi_l(t,r)}{r}\Upsilon_{lm}(\theta,\phi)
\label{nlptwoeq6}
\ee
the eigenvalue equation
\be 
\Delta_{\theta\phi}\Upsilon_{lm}(\theta,\phi)=-l(l+1)\Upsilon_{lm}(\theta,\phi)
\label{nlptwoeq7}
\ee
and transforming to tortoise coordinates defined as (see e.g.  \cite{Cardoso:2003sw,Choudhury:2004ph,Toshmatov:2017qrq})
\be
dr_*\equiv\frac{dr}{f(r)}
\label{nlptwoeq5}
\ee
the KG equation reduces to
\be 
\left(\frac{\partial^2}{\partial t^2}-\frac{\partial^2}{\partial r_*^2}+V_l(r)\right)\Psi_l(t,r_*)=0
\label{nlptwoeq8}
\ee
where $V_l(r)$ is a Regge-Wheeler type potential which when expressed in the original radial coordinate is of the form 
\be
V_l(r)=f(r)\left(\frac{l(l+1)}{r^2}+\frac{f'(r)}{r}(1-s)+m^2\right)
\label{nlptwoeq9} 
\ee
with $s=0$ (spin of the considered field) for the case of a scalar field. This type of effective potential was first derived for “axial” (vector type) perturbations in the Schwarzschild background by Regge-Wheeler \cite{Regge:1957td}. For “polar”(scalar type) gravitational perturbations the effective potential was first derived by Zerilli \cite{Zerilli:1971wd,Zerilli:1970se}. As discussed in \cite{Nashed:2019tuk}, the Regge-Wheeler-Zerilli formalism is based on the assumption of spherical symmetry.

For the solution of Eq. (\ref{nlptwoeq8})  we need to express the Regge-Wheeler potential $V_l(r)$ in tortoise coordinates $ V_{*l}(r_*)\equiv V_l(r(r_*))$. Thus we need to evaluate the integral 
\be 
r_*\equiv \int \frac{dr}{f(r)}=\int\frac{dr}{\sqrt{1-\frac{2M}{r}-\frac{\Lambda}{3}r^2}}
\label{nlptwoeq10}
\ee
To evaluate the integral (\ref{nlptwoeq10})  we follow \cite{Bhattacharya:2018ltm} (see also \cite{Stuchlik:2009jv}) and factorize $f(r)$. Let
\be
\xi=9M^2 \Lambda
\label{xidef}
\ee
For $\xi<1$ there are three real solutions of $f(r)=0$. Two of them correspond to the event and cosmological  horizons ($r_H$ and $r_C$) while  the third is negative ($r_N$) and does not correspond to a physical horizon. The three horizon radii are \cite{1983BAICz..34..129S,Gibbons:1977mu,Guven:1990ubi,Ashtekar:2003hk,Choudhury:2004ph,Stuchlik:2009jv,Toshmatov:2017qrq,Bhattacharya:2018ltm}
\ba
r_H &=& \frac{2}{\sqrt{\Lambda}}\cos\left[\frac{1}{3}\cos^{-1}(3M\sqrt{\Lambda})+\frac{\pi}{3}\right]
\label{nlptwoeq12} \\
r_C  &=& \frac{2}{\sqrt{\Lambda}}\cos\left[\frac{1}{3}\cos^{-1}(3M\sqrt{\Lambda})-\frac{\pi}{3}\right] 
\label{nlptwoeq13} \\
r_N &=& -(r_H+r_C)
\label{nlptwoeq14}
\ea
For $\xi=1$ which corresponds  to the Nariai solution \cite{1950SRToh..34..160N,Bousso:1997wi}) we have an extremal SdS spacetime \cite{Lake:1977ui,Romans:1991nq,1983BAICz..34..129S,Cardoso:2003sw}
\be
r_H=r_C=\frac{2}{\sqrt{\Lambda}}\cos\frac{\pi}{3}=\frac{1}{\sqrt{\Lambda}}\simeq 10^{26}m
\label{nlptwoeq15}
\ee
where in the last equality we have assumed the observed value of $\Lambda=3H_0^2 \Omega_\Lambda$. 
The surface gravity of the SdS metric at a coordinate radius $r_0$ is defined as \cite{Hawking:1974sw,Cardoso:2003sw,Choudhury:2003wd,Choudhury:2004ph}
\be
\kappa_0\equiv \frac{1}{2}\frac{df}{dr}|_{r=r_0}=\frac{M}{r_0^2}-\frac{1}{3}\Lambda r_0
\label{nlptwoeq16}
\ee
and describes the gravitational acceleration of a test particle at position $r_0$. Using Eqs. (\ref{nlptwoeq12}), (\ref{nlptwoeq13}) and (\ref{nlptwoeq14})  to factorize $f(r)$ in Eq. (\ref{nlptwoeq10})  and the definition (\ref{nlptwoeq16}) we may obtain $r_*(r)$ as \cite{Choudhury:2003wd,Bhattacharya:2018ltm}
\be
\begin{split}
r_*=&\int\frac{dr}{\sqrt{1-\frac{2M}{r}-\frac{\Lambda}{3}r^2}}\\
=&\frac{1}{2\kappa_H}\ln\left(\frac{r}{r_H}-1\right)+\frac{1}{2\kappa_C}\ln\left(1-\frac{r}{r_C}\right)+\\
&+\frac{1}{2\kappa_N}\ln\left(1-\frac{r}{r_N}\right)
\end{split}
\label{nlptwoeq17}
\ee
where we note that $\kappa_C$ is negative.

Using now Eqs. (\ref{nlptwoeq9}) and (\ref{nlptwoeq17}) it is easy to make a parametric plot of $V_{*l}(r_*)$ by plotting pairs of $(r_*(r),V_l(r))$ for $r\in [r_H,r_C]$.

From Eq. (\ref{nlptwoeq17}) it is clear that the tortoise coordinates map the event and cosmological horizons to $\pm \infty$
\be
\begin{split}
r\rightarrow r_H\Longrightarrow r_*\rightarrow - \infty\\
r\rightarrow r_C\Longrightarrow r_*\rightarrow + \infty\\
\end{split}
\label{nlptwoeq18}
\ee
The Regge-Wheeler potential $V_{*l}(r_*)$ of Eq. (\ref{nlptwoeq9}) has the important property that it vanishes at both infinities ($\pm \infty$). This is easy to see since
\be
\begin{split}
&V(r_H)=V(r_C)=0\Longrightarrow\\
&V_*(r_*\rightarrow - \infty)=V_*(r_*\rightarrow + \infty)=0
\end{split}
\label{nlptwoeq19}
\ee
As shown below, this property leads to a simple asymptotic solution of Eq. (\ref{nlptwoeq8}). 

At this point we introduce a rescaling of the radial and time coordinates by $M$ ($r/M \rightarrow  \bar r$, $t/M \rightarrow  \bar t$) and use the dimensionless parameters $\xi$ (defined in Eq. (\ref{xidef})) and 
\be
m\; M\equiv \frac{GM m}{\hbar c}
\label{mMdef}
\ee
In order to search for scalar field instabilities we also use the following ansatz in Eq. (\ref{nlptwoeq8})
\be
\Psi_l(t,r_*)=\left( C_1 e^{\Omega t}+C_2 e^{-\Omega t}\right)u_l(r_*)
\label{nlptwoeq20}
\ee
This ansatz along with the above rescaling transforms Eq. (\ref{nlptwoeq8}) to a Schrodinger-like Regge-Wheeler equation of the form
\be
\frac{du_l^2}{dr_*^2}-M^2 \left(\Omega^2+V_{*l}(r_*)\right)u_l(r_*)=0
\label{nlptwoeq21}
\ee
where  $r_*\in (-\infty,+\infty)$ and 
\begin{widetext}
\be
M^2 \; V_{*0}(r(r_*))=\left(1-\frac{2}{r(r_*)}-\frac{1}{27}\xi r(r_*)^2\right)\left(\frac{2}{r(r_*)^3}-\frac{2}{27}\xi+m^2M^2\right)
\label{nlptwoeq59b}
\ee
\begin{figure*}
\begin{centering}
\includegraphics[width=0.97\textwidth]{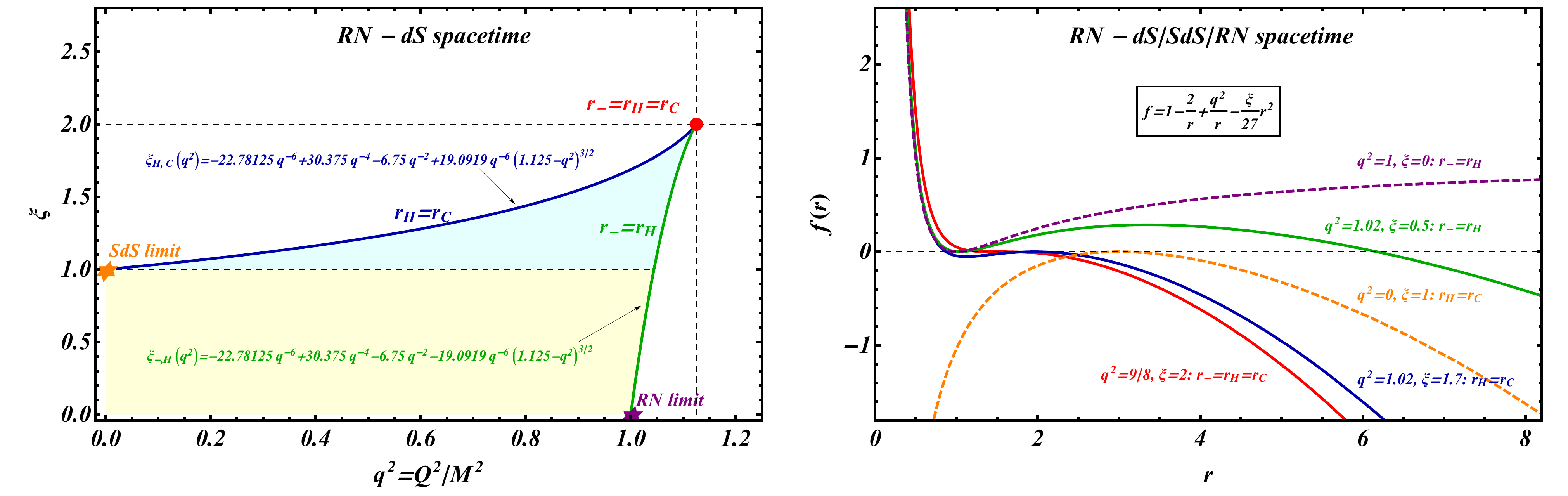}
\par\end{centering}
\caption{The critical values $\xi_{H,C}(q^2)$ (with $0<q^2<9/8$)  and $\xi_{-,H}(q^2)$  (with $1<q^2<9/8$) as a function of  $q^2$  at which $r_H=r_C$ and  $r_-=r_H$ respectively (left panel). The colored shaded regions correspond to the physical corresponding regions of Fig. \ref{figregrrate} discussed below. The metric function $f(r)$  as a function of $r$ in the case of the  RN-dS/SdS/RN spacetimes for critical value $\xi_{H,C}$ (when event and cosmological horizons coincide) and  $\xi_{-,H}$ (when inner Cauchy and outer event horizons coincide) (right panel). The blue, green and red solid curves correspond to RN-dS spacetime while the purple and orange dashed curves correspond to RN and SdS spacetime respectively.} 
\label{figximaxfr} 
\end{figure*}
\end{widetext}
In (\ref{nlptwoeq21}), (\ref{nlptwoeq59b}) we have omitted the bar of the rescaled coordinates and in  (\ref{nlptwoeq59b}) we have fixed $l=0$. Since $V_{*l}(r)>V_{*l=0}$, the most unstable scales are the large angular scales $l=0$. This behavior is similar to the case of the Minkowski spacetime discussed in the introduction where the scale corresponding to $k=0$ was the most unstable scale (largest growth rate, smallest lifetime). Thus in what follows we focus on the $l=0$ modes. If these modes are stable then all scales ($l>0$) are also stable.
 
\subsection{Reissner-Nordström-deSitter background}
We now generalize the metric of the previous section by including charge in the black hole metric. The RN-dS spacetime  is defined by  the metric function \cite{Romans:1991nq,Brady:1996za} 
\be
\begin{split}
f(r)=1-\frac{2M}{r}+\frac{Q^2}{r^2}-\frac{\Lambda}{3}r^2=\\
=1 - \frac{2}{r} + \frac{q^2}{r^2} - \frac{\xi}{27}r^2
\end{split}
\label{rndsf}
\ee
where  $\xi$  is defined in Eq. (\ref{xidef}),  $q\equiv \frac{Q}{M}$ (where $Q$ is the black hole electric charge) and in the second equality we have used the rescaling $r/M\rightarrow r$.

The horizons are obtained by solving the equation  $f(r)=0$. For $\xi<2$ and $q^2<9/8$ there are four real solutions \cite{Romans:1991nq}. Two of them  correspond to the inner (Cauchy) and outer (event) horizons of a RN black hole $r_-$ and $r_+=r_H$ (with $0< r_-< r_H$) respectively. The third corresponds to the cosmological horizon $r_C$ (with $r_C>r_H$) while  the fourth $r_N$ (with $r_N=-(r_-+r_H+r_C)$) is negative  and does not correspond to a physical horizon. 

The three horizons coincide at \cite{Romans:1991nq}
\be
r_-=r_H=r_C=\frac{3}{\sqrt{2\xi}} 
\label{rndsexthor}
\ee
when $\xi=2$ and $q^2=9/8$. 

By demanding that two of the physical horizons coincide we set the discriminant of the quartic equation $f(r)=0$ to zero and obtain the equation \cite{Romans:1991nq,Montero:2019ekk})
\be 
1-q^2-\xi+\frac{4}{3}\xi q^2-\frac{8}{27}\xi q^4-\frac{16}{729}\xi^2 q^6=0
\label{discreq}
\ee
which has real solutions for $\xi$ when $0<q^2<\frac{9}{8}$.  The  critical value  $\xi_{H,C}$ at which $r_H=r_C$ and the corresponding value $\xi_{-,H}$ at which $r_-=r_H$ may be obtained in terms of $q^2$ by solving Eq. (\ref{discreq}) as
\begin{widetext}
\be 
\xi_{H,C}=-22.7813 q^{-6}+30.375 q^{-4}-6.75 q^{-2}+ 19.0919 q^{-6} (1.125-q^2)^{\frac{3}{2}}
\ee
\be 
\xi_{-,H}=-22.7813 q^{-6}+30.375 q^{-4}-6.75 q^{-2}- 19.0919 q^{-6} (1.125-q^2)^{\frac{3}{2}}
\ee
\end{widetext}
The first case corresponds to the charged Nariai solution \cite{Montero:2019ekk}. The critical values $\xi_{H,C}(q^2)$  and $\xi_{-,H}(q^2)$  as a function of  $q^2$  are shown in Fig. \ref{figximaxfr} (left panel). The critical value $\xi_{H,C}(q^2)$ that leads to a coincidence between the event and cosmological horizons (blue line) varies between $1$ (SdS limit, $q=0$) and $2$ (triple horizon coincidence limit, $q^2=9/8$). The corresponding form of the function $f(r)$ in these (and in other) limits is shown in Fig. \ref{figximaxfr} (right panel). The orange line corresponds to the coincidence of the event with the cosmological horizon $r_H=r_C$ in the SdS limit while the blue line shows the coincidence of the same roots of $f(r)$ ($r_H=r_C$) in the general RN-dS case with $q^2=1.02$. In both cases the local maximum of $f(r)$ occures at $f(r)=0$.

In the case of RN-dS, we study tachyonic instabilities of the neutral massive scalar field perturbations in the event-cosmological horizon region, defined as $r_+=r_H<r<r_C$ using tortoise coordinates $r_*(r)$ defined as
\begin{widetext}
\be
r_*=\int\frac{dr}{\sqrt{1-\frac{2M}{r}+\frac{Q^2}{r^2}-\frac{\Lambda}{3}r^2}}
=\frac{1}{2\kappa_-}\ln\left(\frac{r}{r_-}-1\right)+\frac{1}{2\kappa_H}\ln\left(\frac{r}{r_H}-1\right)
+\frac{1}{2\kappa_C}\ln\left(1-\frac{r}{r_C}\right)+\frac{1}{2\kappa_N}\ln\left(1-\frac{r}{r_N}\right)
\label{rndstort}
\ee
\end{widetext} 
with $\kappa_i$ ($i=-,H,C$) the surface gravity for the   horizon $r=r_i$
\be
\kappa_i\equiv \frac{1}{2}\frac{df}{dr}|_{r=r_i}=\frac{M}{r_i^2}-\frac{Q^2}{r_i^3}-\frac{1}{3}\Lambda r_i
\label{rndssurgr}
\ee
where we note that $\kappa_-<0 $ and $\kappa_C<0$.
It is easy to see that the tortoise coordinates $r_*(r)$ shift the horizons $r_H$ and $r_C$  to $\pm \infty$. 

The values of the inner (Cauchy) and outer (event) horizon in the case of RN background ($\Lambda=0$) for $Q<M$ are (see e.g. \cite{Misner:1974qy})
\be 
 r_{\pm}=M\pm\sqrt{M^2-Q^2}
\label{rnhor}
\ee 
In the case of RN-dS spacetime a rescaling of the radial and time coordinates by $M$ ($ r/M\rightarrow r$, $t/M\rightarrow t$) and the introduction of the dimensionless parameters $\xi$ (defined in Eq. (\ref{xidef})), $q=Q/M$ and $mM$ (defined in Eq. (\ref{mMdef})) lead to the Schrodinger-like equation (\ref{nlptwoeq21}) with maximum scale ($l=0$) generalized Regge-Wheeler potential of the form 
\begin{widetext}
\be
M^2 \; V_{*0}(r(r_*))=\left(1-\frac{2}{r(r_*)}+\frac{q^2}{r(r_*)^2}-\frac{1}{27}\xi r(r_*)^2\right)\left(\frac{2}{r(r_*)^3}-\frac{2q^2}{r(r_*)^4}-\frac{2}{27}\xi+m^2M^2\right)
\label{rndspot}
\ee
\end{widetext}
with $r_*\in (-\infty,+\infty)$.
\begin{figure*}
\begin{centering}
\includegraphics[width=0.97\textwidth]{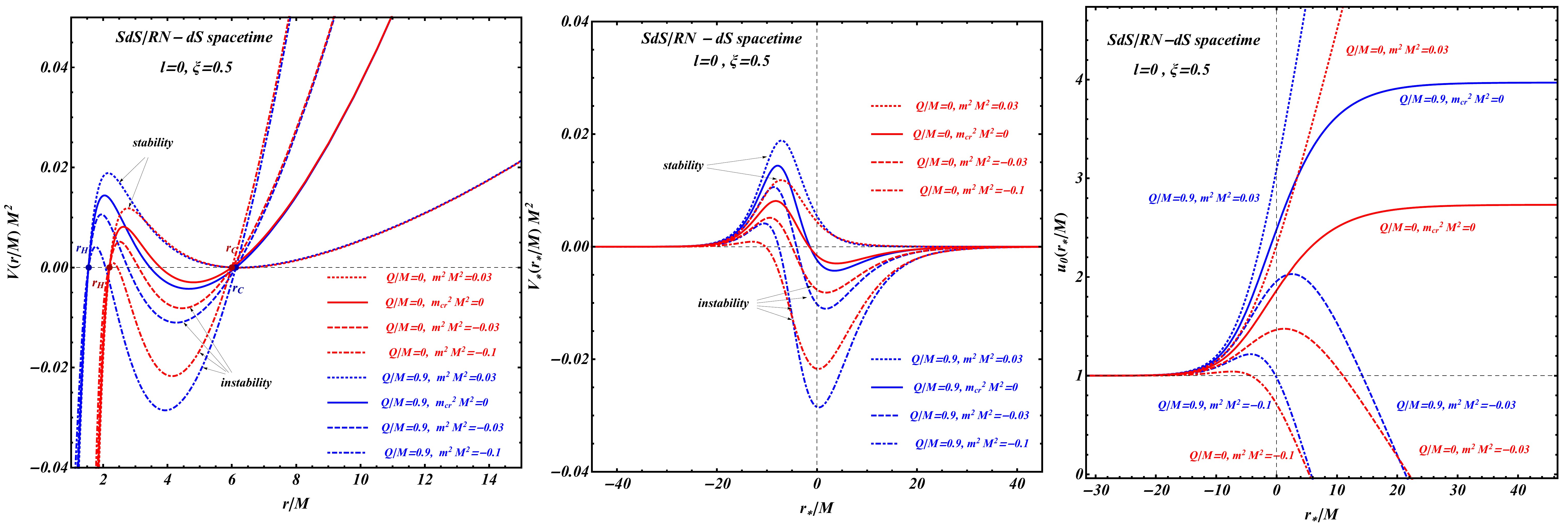}
\par\end{centering}
\caption{The  $m^2 M^2$ dependent Regge-Wheeler dimensionless potentials  $V M^2$ (left panel) and  $V_* M^2$  (middle panel) as a function of  $r/M$ and  $r_*/M$ respectively in the cases of the SdS ($Q=0$) (red curves) and RN-dS ($Q/M=0.9$) (blue curves) spacetimes for angular scale $l=0$ and  dimensionless parameter fixed to $\xi=0.5$. The solid curves correspond to  the critical value of the scalar field mass $m_{cr}^2 M^2=0$.  The right panel demonstrates the process for identifying the zero eigenvalue eigenstate i.e. setting $\Omega=0$ in Eq. (\ref{nlptwoeq21}) and increasing the dimensionless parameter $m^2 M^2$ until the solution $u_0(r_*/M)$ satisfies both end boundary conditions (\ref{nlptwoeq29})-(\ref{nlptwoeq32}) for $\Omega=0$. This value of $m^2 M^2$ is the critical value for the considered value of $\xi$. The potential gets deeper and more accepting to bound states (instabilities) as the $m^2 M^2$ gets lower.} 
\label{fig1} 
\end{figure*}
\begin{figure*}
\begin{centering}
\includegraphics[width=0.97\textwidth]{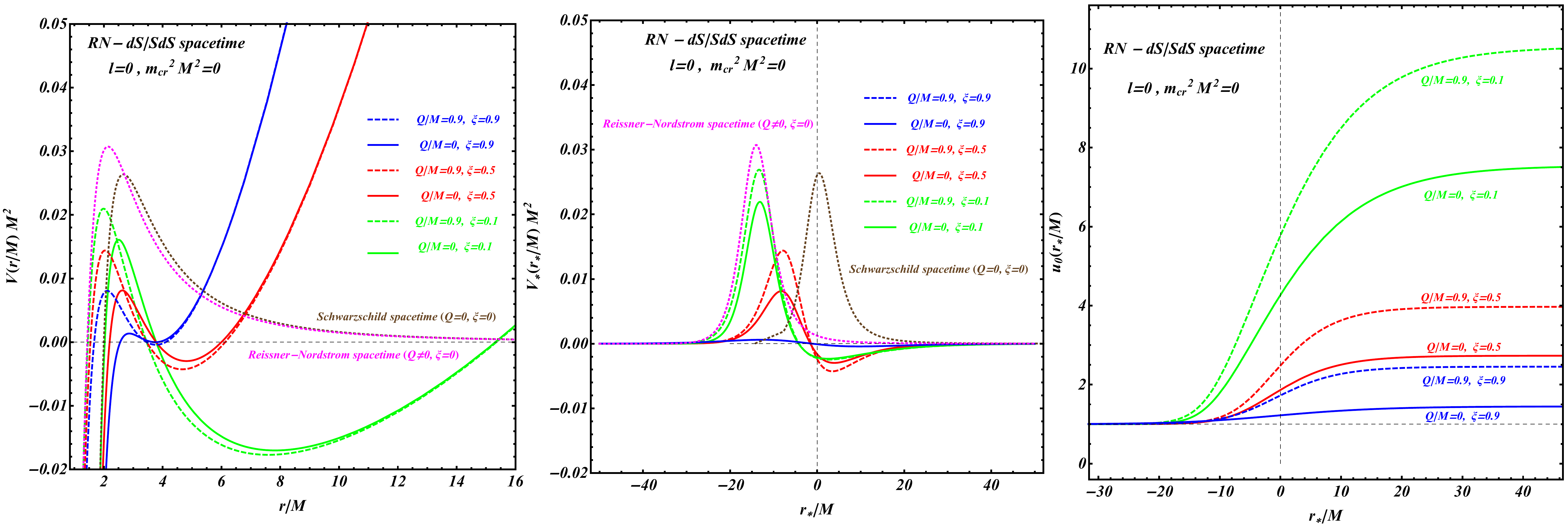}
\par\end{centering}
\caption{The  $\xi$ dependent Regge-Wheeler dimensionless potentials  $V M^2$ (left panel) and  $V_* M^2$  (middle panel) as a function of  $r/M$ and  $r_*/M$ respectively in the case of the SdS (solid curves) and RN-dS (dashed curves) spacetimes for angular scale $l=0$ and critical value for $m^2=m_{cr}^2=0$. The radial function $u_0(r_*/M)$ (right panel) which is the radial zero mode solution of Schrodinger like equation (\ref{nlptwoeq21}) with $\Omega=0$ and boundary conditions (\ref{nlptwoeq29}) and (\ref{nlptwoeq30}) at large negative $r_*$. For critical value of the scalar field mass $m_{cr}^2M^2=0$ the  boundary conditions (\ref{nlptwoeq31}) and (\ref{nlptwoeq32}) at large positive $r_*$ are satisfied. The  brown and purple dotted curves correspond to the pure  Schwarzschild ($Q=0$, $\xi=0$) and RN ($Q\neq0$, $\xi=0$) backgrounds respectively. The potential gets deeper as $\xi$ decreases and $Q/M$ increases. However, since the local maximum of the potential also increases as the potential gets deeper, the critical value $m_{cr}M$ for the existence of bound states remains the same and equal to zero in all cases.} 
\label{fig2} 
\end{figure*}

\section{Numerical solution: Parameter region for instability, Growth rate.}  
\label{numsol} 

The questions we want to address in this section are the following: 
\begin{itemize}
    \item 
    {\it What is the critical value $m_{cr}(q,\xi)^2$ such that for $m^2>m_{cr}^2$ Eq. (\ref{nlptwoeq21})  with a real $\Omega^2$ has no bound state solutions (no instabilities) respecting the physically acceptable boundary conditions that correspond to finite field values at the two horizons ($r_* \rightarrow \pm \infty$)?}
    \item
    {\it What is the growth rate $\Omega(q,\xi,m^2M^2)$ of tachyonic instabilities ($m^2<m_{cr}^2$) and how does this growth rate compare with the corresponding growth rate in a flat Minkowski spacetime?}
\end{itemize}
We thus solve the Schrodinger-like Regge-Wheeler equation (\ref{nlptwoeq21}) and for fixed values of $q$ and $\xi$ we start from a low negative $m^2$ and identify the ground state solution. Then we increase the value of $m^2$ until there are no bound states (instability modes) with physically acceptable boundary conditions. At the critical value $m^2=m_{cr}^2$ there will only be a zero mode solution with eigenvalue $\Omega=0$ (infinite lifetime mode). Such a mode may be interpreted as a scalar hair zero mode. As discussed in the 'Introduction', in Minkowski space ($M=\Lambda=Q=0$), we have $m_{cr}=0$. {\it Does this value of $m_{cr}$ change in RN-dS or in SdS spacetimes?}

To address this question we must first find the required `physical boundary conditions'. We demand that the physically acceptable solution should be finite on the two horizons i.e. 
\be
\begin{split}
u_0(r_*\rightarrow \infty)<+\infty\\
u_0(r_*\rightarrow -\infty)<+\infty\\
\end{split}
\label{nlptwoeq23}
\ee
Since $V_{*0}(r_*)$ goes exponentially fast to 0 for $r_*\rightarrow \pm \infty$, we conclude that the general asymptotic solution of Eq. (\ref{nlptwoeq21})  is
\be
u_0(r_*\rightarrow\pm\infty)=A e^{\Omega r_*}+B e^{-\Omega r_*}
\label{nlptwoeq24}
\ee
For finiteness we demand
\ba
u_0(r_*\rightarrow +\infty)&=& B e^{-\Omega r_*}
\label{nlptwoeq25}\\
u_0(r_*\rightarrow -\infty)&=&A e^{\Omega r_*}
\label{nlptwoeq26}
\ea
These imply 
\ba
u_0'(r_*\rightarrow +\infty)&=&-\Omega B e^{-\Omega r_*} 
\label{nlptwoeq27}\\
u_0'(r_*\rightarrow -\infty)&=&\Omega A e^{\Omega r_*}
\label{nlptwoeq28}
\ea
where we can rescale $u_0(r_*)$ such that $A=1$. These boundary conditions leading to instability may be associated with bound states ($\Omega^2>0$, $\Omega \in \mathbb{R}$) of the Schrodinger-like equation (\ref{nlptwoeq21}) with effective Regge-Wheeler potential $V_{*0}(r_*)$ (see Eq. (\ref{nlptwoeq59b}) for SdS spacetime and Eq. (\ref{rndspot}) for RN-dS spacetime). Our search for scalar instabilities ($\Omega^2>0$)  should be contrasted with the search for the values of QNMs which involves propagating boundary conditions at the horizons. These studies have also indicated the presence of scalar instabilities in a different physical setup (charged massive scalar field in Kerr-Newman black holes with positive $m^2$  \cite{Konoplya:2013rxa}).

\begin{figure}
\begin{centering}
\includegraphics[width=0.47\textwidth]{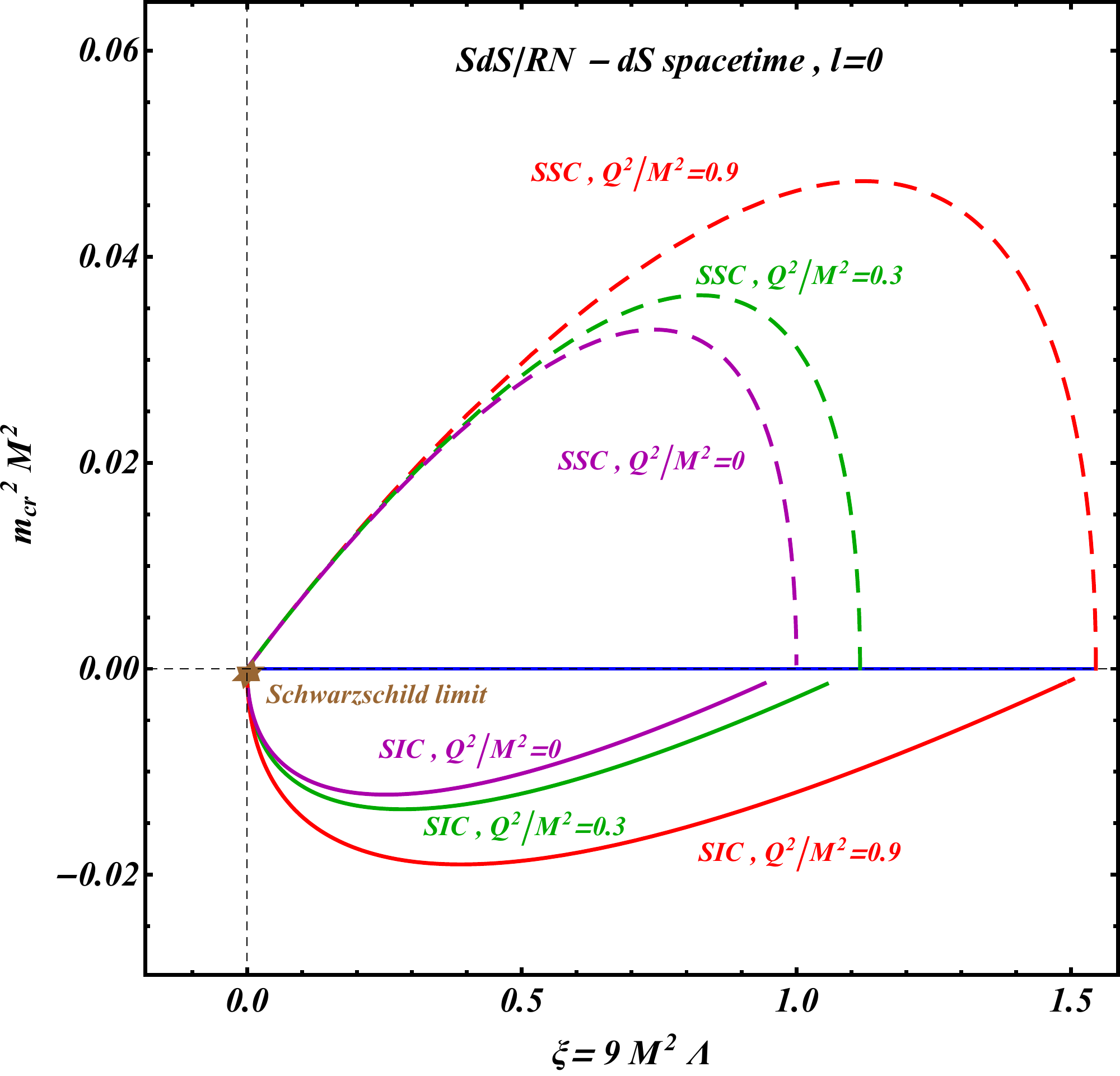}
\par\end{centering}
\caption{The critical value of the scalar field mass $m_{cr}^2M^2$ is zero and independent of the dimensionless parameter  $\xi$ (with $\xi\in [0,\xi_{H,C}(q)]$) in the case of the SdS and RN-dS spacetime  (blue straight line) for $l=0$.  The solid curves show  the form of $m_{cr}(q,\xi)^2M^2$ that saturates the Sufficient for Instability Criterion (SIC) Eq. (\ref{nlptwoeq33}) while the corresponding dashed curves shows the forms of $m_{cr}(q,\xi)^2M^2$ that saturate the Sufficient for Stability Criterion (SSC) Eq. (\ref{nlptwoeq34}) for three values of $Q/M$. As expected, the exact value of $m_{cr}M=0$ is between the SIC lines (lower lines) and SSC lines (upper lines) so that none of the criteria is violated (SSC or SIC).} 
\label{fig4new} 
\end{figure}
  
\begin{figure}
\begin{centering}
\includegraphics[width=0.47\textwidth]{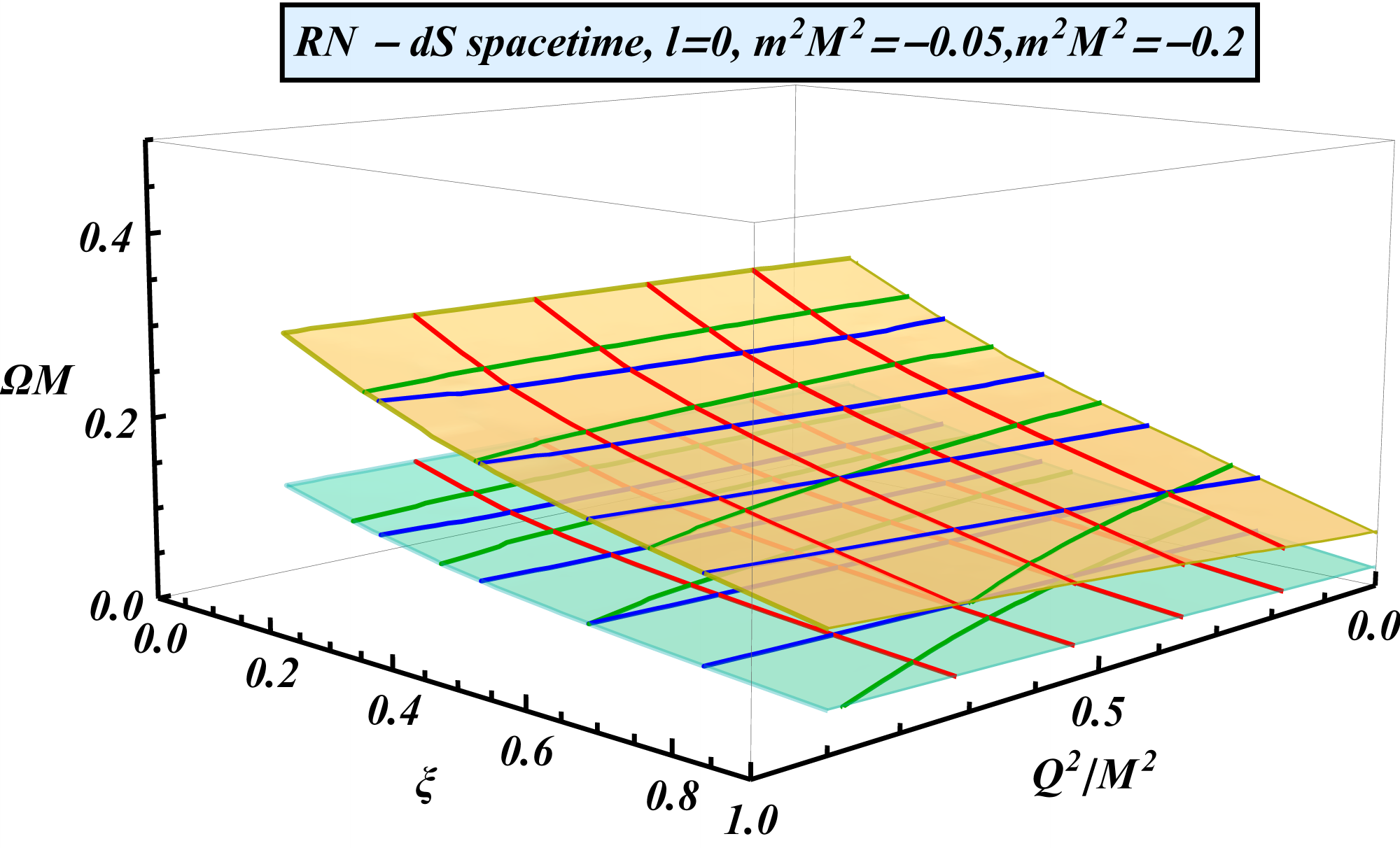}
\par\end{centering}
\caption{The dimensionless growth rate  of the instability $\Omega M$ as a function of the dimensionless parameters $\xi$ and $q^2=Q^2/M^2$ for scalar field mass $m^2M^2=-0.05$ (cyan surface) and $m^2M^2=-0.2$ (yellow surface). } 
\label{fig3dim} 
\end{figure}

The Regge-Wheeler potential $V_{*0}(r_*)$ is mostly accepting bound states for lower values of $m^2M^2$ and for higher values of $Q/M$. This is demonstrated in Fig. \ref{fig1} where we show the form of $V_{*0}(r_*)$ for various values of the dimesionless parameter $m^2M^2$ in the cases of the SdS ($Q=0$) and RN-dS ($Q/M=0.9$) spacetimes for angular scale $l=0$ and  $\xi=0.5$ indicating that as $m^2M^2$ gets lower and as $Q/M$ gets higher, the minimum of the Regge-Wheeler potential gets deeper and thus it becomes more accepting to the existence of bound states (instabilities). The critical value $m_{cr}(q,\xi)^2$ is such that for $m^2>m_{cr}^2$ there are no bound states (instabilities) respecting the boundary conditions (\ref{nlptwoeq25}),  (\ref{nlptwoeq26}),  (\ref{nlptwoeq27}) and  (\ref{nlptwoeq28}). 

The critical value $m_{cr}(q,\xi)^2$ is obtained by solving Eq.  (\ref{nlptwoeq21})  with boundary conditions (\ref{nlptwoeq25}),  (\ref{nlptwoeq26}),  (\ref{nlptwoeq27}) and  (\ref{nlptwoeq28})  for a zero eigenvalue $\Omega=0$ corresponding to a borderline unstable mode (zero mode) with infinite lifetime and zero growth rate. For such a zero mode, the boundary conditions (\ref{nlptwoeq25}),  (\ref{nlptwoeq26}),  (\ref{nlptwoeq27})  and  (\ref{nlptwoeq28})  become
\ba
u_0(r_*\rightarrow -\infty)&=&1
\label{nlptwoeq29}\\
u_0'(r_*\rightarrow -\infty)&=&0
\label{nlptwoeq30}\\
u_0(r_*\rightarrow +\infty)&=&B
\label{nlptwoeq31}\\
u_0'(r_*\rightarrow +\infty)&=&0
\label{nlptwoeq32}
\ea
where we have set $A=1$. 

In practice we use the shooting method in solving Eq. (\ref{nlptwoeq21})  with $\Omega=0$, fixed $\xi$, $q$, boundary conditions (\ref{nlptwoeq29}), (\ref{nlptwoeq30}) at large negative $r_*$ and adjust $m^2M^2$ until the boundary conditions (\ref{nlptwoeq31}) and (\ref{nlptwoeq32}) are satisfied (see Fig. \ref{fig1} right panel). By repeating this process for several values of  $q^2 \in [0,\frac{9}{8}]$ and $\xi\in [0,\xi_{H,C}(q)]$  we have found $m_{cr}(\xi,q)^2=0$ i.e. the zero mode appears at $m^2=0$ for all parameter values $\xi, q$ where there is a finite distance between the event and the cosmological horizons. 

In Fig. \ref{fig2} we show  the form of the Regge-Wheeler potentials  $V_0(r/M)$ and $V_{*0}(r_*/M)$ as well as  the radial zero mode solution $u_0(r_*/M)$ for the critical value $m_{cr}(q,\xi)=0$ for $\xi=0.1,0.5,0.9$ in the case of the SdS spacetime ($q=0$) and in the case of RN-dS spacetime ($q=0.9$). Notice that in the absence of a cosmological horizon  ($\xi=0$, pure Schwarzschild and Reissner-Nordström spacetimes) the Regge-Wheeler potential $V_*$ is positive everywhere for $m=0$ and the absence of bound states is obvious. However, this is not the case for $\xi >0$ which requires numerical solution of the Schrodinger-like equation for the determination of $m_{cr}$.

There is a simple semi-analytical way to derive sufficient conditions for instability and for stability and thus test the validity of the numerically obtained form of $m_{cr}^2=0$ for various values of the parameters $\xi$ and $q$.  It is well known that a sufficient condition for the existence of bound states in a Schrodinger equation potential $V_{*0}(r_*)$ is the following Sufficient for Instability Criterion (SIC) \cite{doi:10.1119/1.17935,Dotti:2004sh,Myung:2018vug}
\be 
\begin{split}
&I_{SIC}=\int_{-\infty}^{+\infty}V_{*0}(r_*)dr_*<0\Longrightarrow\\
&\int_{r_H}^{r_C} \frac{V_0(r)}{f(r)}dr=\int_{r_H}^{r_C}\left(\frac{l(l+1)}{r^2}+\frac{f'(r)}{r}+m^2\right)_{l=0}dr=\\
&\int_{r_H}^{r_C}\left(\frac{l(l+1)}{r^2}+\frac{2}{r^3}-\frac{2q^2}{r^4}-\frac{2}{27}\xi+m^2M^2\right)_{l=0}dr<0
\end{split}
\label{nlptwoeq33}
\ee
where we have used Eqs. (\ref{nlptwoeq9}) and  (\ref{rndspot}) for the form of $V_0(r)$ and the dimensionless parameters $\xi$ and $q$. In addition, a positive definite potential can not have bound states (negative eigenvalues corresponding to $\Omega^2>0$) and thus in such a potential we would only have stable oscillating modes ($\Omega^2<0$). Thus a Sufficient for Stability Criterion (SSC) is that the minimum of the Schrodinger potential should be positive i.e.
\be
V_{0 min}(r_{min})>0
\label{nlptwoeq34}
\ee
Using the SIC and the SSC we have constructed the upper and lower curves in Fig. \ref{fig4new} which correspond to the values of $m(\xi)^2M^2$ that saturate the SSC (upper curves) and SIC (lower curves). Also, using the SSC we find an analytical expression $m^2(\xi) M^2$ for $Q=0$ (upper curve in Fig. \ref{fig4new} see Appendix \ref{Appendix}). Thus, by construction all parameter values below the lower curves satisfy the SIC Eq. (\ref{nlptwoeq33})  and must correspond to tachyonic instabilities while all parameter values above the upper curves of Fig.  \ref{fig4new} satisfy the SSC Eq. (\ref{nlptwoeq34})  and have no instabilities. As expected the precise numerically obtained values of $m_{cr}(\xi)^2=0$ are between the SIC and SSC curves so that none of the two sufficient (but not necessary) conditions is violated. 
\begin{figure*}
\begin{centering}
\includegraphics[width=0.97\textwidth]{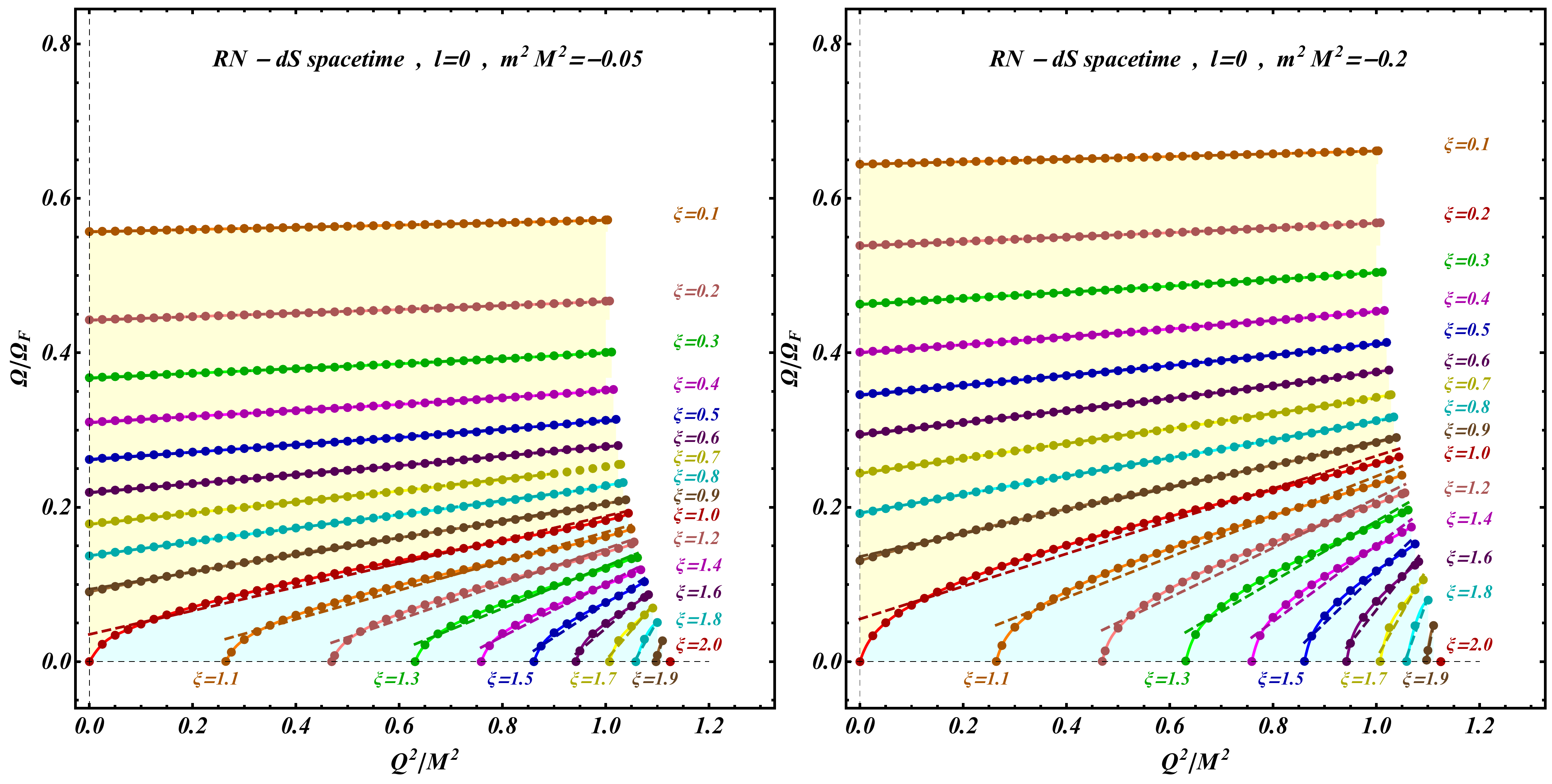}
\par\end{centering}
\caption{The $\xi$ dependent relative growth rate  of the instability $\Omega/\Omega_F$ (with $\Omega_F$ the growth rate  of the instability in flat spacetime) as  a function of the dimensionless parameter $q^2=Q^2/M^2$ for the scalar field mass $m^2M^2=-0.05$ (left panel) and $m^2M^2=-0.2$ (right panel). The curves for a given parameter value $\xi$ (with  $\xi<1$)  turn out to be straight lines. The range of values of $\xi$ and $q$ is determined by the physically interesting parameter region between the green and blue lines of Fig. \ref{figximaxfr}.The parameter region corresponding to linear behavior of $\Omega(q^2)$ (yellow region) is also shown in Fig. \ref{figximaxfr}. } 
\label{figregrrate} 
\end{figure*}
Even though the value $m_{cr}(q,\xi)=0$ for the emergence of tachyonic instabilities is independent of the metric parameters and remains the same in the RN-dS spacetime as in the flat Minkowski spacetime, the growth rate $\Omega(q,\xi,m)$ of tachyonic instabilities ($m^2<0$) does have a dependence on the metric parameters. In order to identify this dependence we consider an unstable mode with fixed  $m^2<m_{cr}(q,\xi)^2=0$ and given $\xi$ and $q$,  we find the growth rate $\Omega$ of the instability by finding the ground state eigenvalue\footnote{Possible excited states would correspond to lower values of $\Omega$ and thus lower growth rate. We thus find the maximum possible growth rate of instabilities for a given set of parameters.}  $\Omega^2$ and eigenfunction $u_0(r_*)$ of the Schrodinger-like equation (\ref{nlptwoeq21}) which has no nodes and obeys the boundary conditions (\ref{nlptwoeq26})-(\ref{nlptwoeq28}), (\ref{nlptwoeq25}) and  (\ref{nlptwoeq27}).  We thus construct Fig. \ref{fig3dim} which shows the dimensionless growth rate  of the instability $\Omega M$ as a function of the dimensionless parameters $\xi$ and $q^2=Q^2/M^2$ for scalar field mass $m^2M^2=-0.05$ and $m^2M^2=-0.2$. Clearly, when $\xi$ increases and/or $q$ decreases towards $0$, the growth rate  of the instability $\Omega M$ decreases and as $m^2M^2\rightarrow 0$ we have  $\Omega M\rightarrow 0$ (the zero mode is reached). In addition to this interesting monotonic behavior of the instability growth rate $\Omega$ with respect to the metric parameters, $\Omega$ also remains smaller than its flat space value $\Omega_{F}=\vert m \vert$. This is demonstrated in Fig. \ref{figregrrate} where we show the dependence of $\frac{\Omega}{\Omega_F}$ on $q^2$ for various values of $\xi$ for $m^2 M^2=-0.05$ (left panel) and $m^2 M^2=-0.2$ (right panel). We have considered parameter values between the green and blue lines of Fig. \ref{figximaxfr} where three distinct horizon exist in the RN-dS metric. The following observations can be made based on  Figs. \ref{fig3dim}, \ref{figregrrate} 
\begin{itemize}
    \item  The relative growth rate of the tachyonic instabilities $\frac{\Omega}{\Omega_F}$ is a monotonically increasing function of $q^2$ and a monotonically decreasing function of $\xi$.
    \item $\frac{\Omega}{\Omega_F}$  is significantly smaller than unity. This reduction implies that background curvature and especially the combination of an event horizon with a cosmological horizon tend to delay the evolution of instabilities.
    \item There is a linear relation between $\frac{\Omega}{\Omega_F}$  and $q^2$ for fixed $\xi<1$. This is evident in both Fig. \ref{figregrrate} and in Fig. \ref{fig3dim}. For example the straight blue lines of Fig. \ref{fig3dim} correspond to the dependence of $\Omega\; M$ on $q^2$ for fixed $\xi$ which are equivalent to the straight lines of Fig. \ref{figregrrate}. Notice that this linear relation is violated for $\xi>1$ (see shaded regions in Figs \ref{figximaxfr}, \ref{figregrrate}) .
    \item The growth rate $\Omega$ is a decreasing function of $\vert m \vert^2$ which goes to zero as $m^2\rightarrow m_{cr}^2=0$ where the zero mode develops. This is illustrated in more detail in Fig. \ref{figom4q}.  
\end{itemize}
\begin{figure*}
\begin{centering}
\includegraphics[width=0.97\textwidth]{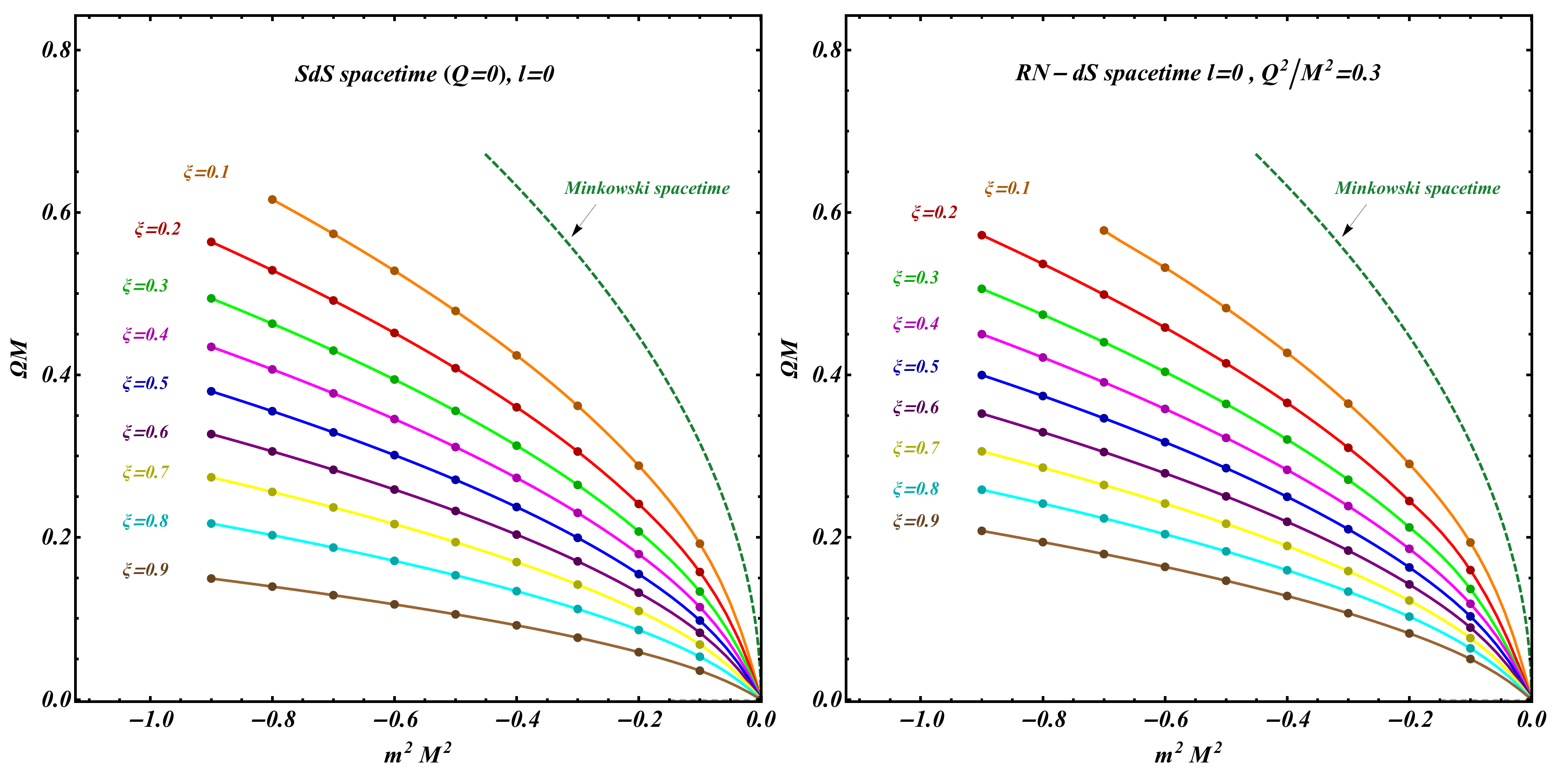}
\par\end{centering}
\caption{The $\xi$ dependent dimensionless growth rate  of the instability $\Omega M$ as a function of  the scalar field mass $m^2M^2$ (with $m(\xi)^2<m_{cr}(\xi)^2=0$) for dimensionless parameters $Q^2/M^2=0$ (SdS spacetime) (left panel) and  $Q^2/M^2=0.3$ (RN-dS spacetime) (right panel).  The green dashed curves correspond to $\Omega M(m^2M^2)$ in the case of the Minkowski spacetime. Clearly, for a given field mass, the growth rate is more suppressed in the absence of charge and for higher values of $\xi$. } \label{figom4q} 
\end{figure*}
The crucial feature of the RN-dS metric that has lead to the above described trend for delay of instability growth of the tachyonic modes is the combination of the cosmological horizon with an event horizon.  This combination, limits the range of negative values of the Regge-Wheeler potential in tortoise coordinates for $m^2<0$ and thus makes it less accepting to bound states and instabilities. In the absence of a cosmological horizon the Regge-Wheeler potential in tortoise coordinates would remain negative out to $r_*\rightarrow \infty$. This is illustrated in the next section.

\section{Limiting cases with a single horizon:  pure desitter and pure Schwarzschild spacetimes} \label{singlehorizon}
We now consider separately the two single horizon limiting cases: pure deSitter and pure Schwarzschild spacetimes in order to isolate the effects of the cosmological and event horizons.
\subsection{Pure deSitter background}
\begin{figure}
\begin{centering}
\includegraphics[width=0.47\textwidth]{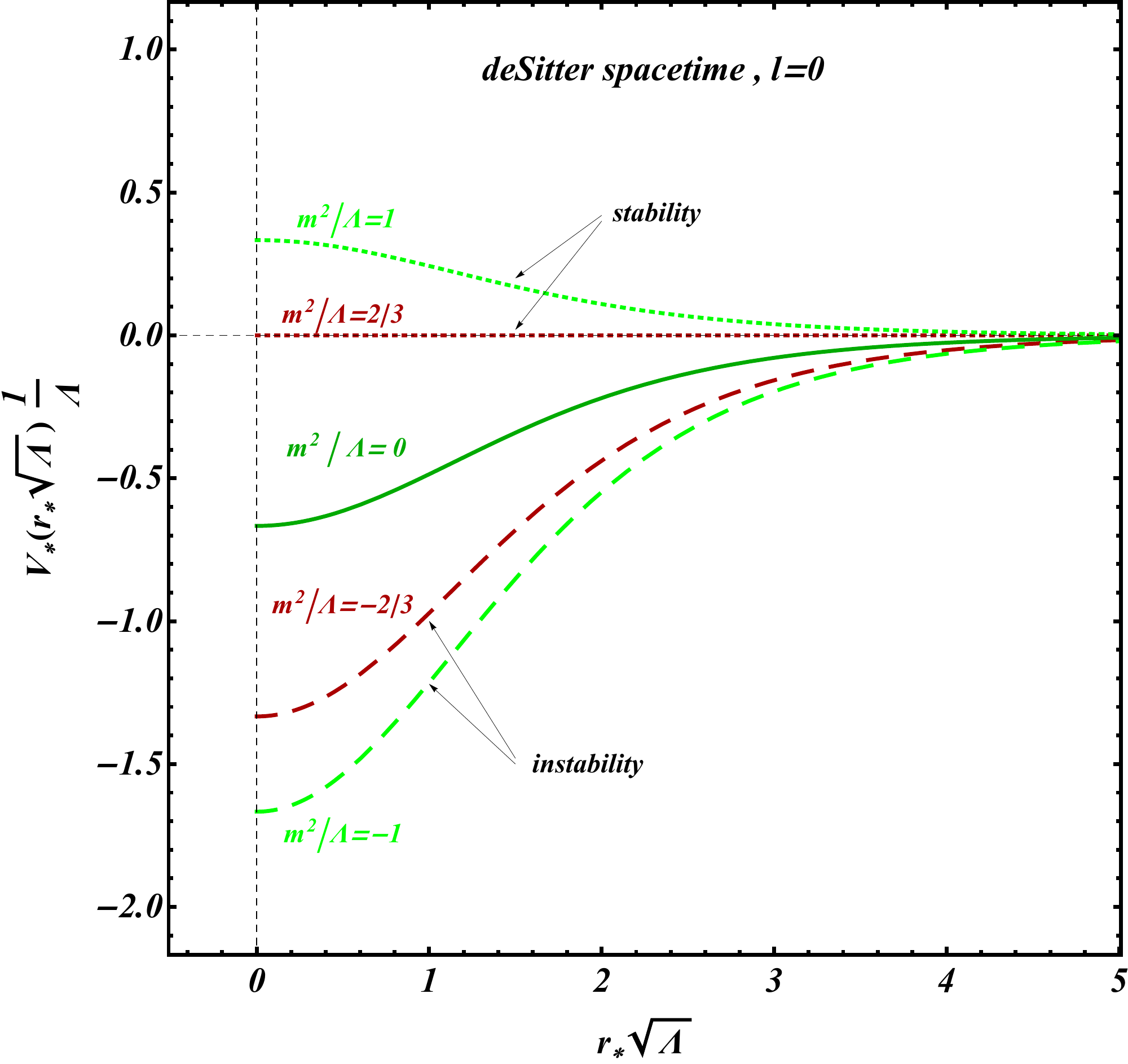}
\par\end{centering}
\caption{The  $m^2/\Lambda$ dependent Regge-Wheeler dimensionless potential $V_*/\Lambda$  as a function of   $r_*\sqrt{\Lambda}$ in the case of the deSitter spacetime ($M=0$,  $\xi=0$) for angular scale $l=0$ . The  green solid curve corresponds to the critical value of the scalar field mass $m_{cr}^2/\Lambda=0$. The dotted ($m^2/\Lambda>0$) and dashed ($m^2/\Lambda<0$) curves correspond to non-existence of bound states (stabilities) and existence of bound states (instabilities) respectively.} 
\label{splvrtds} 
\end{figure}

\begin{figure}
\begin{centering}
\includegraphics[width=0.47\textwidth]{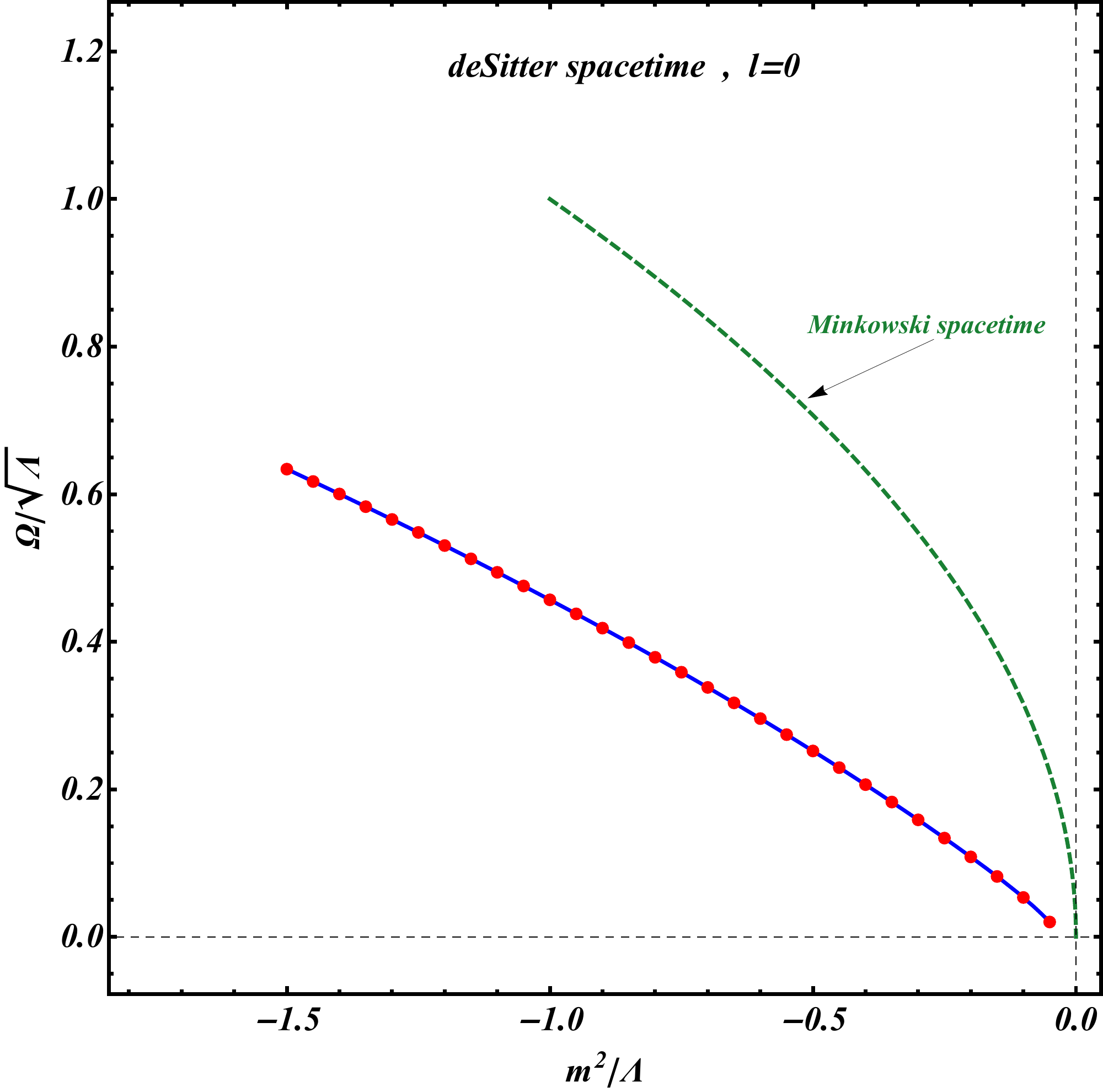}
\par\end{centering}
\caption{The dimensionless growth rate  of the instability $\Omega/\sqrt{\Lambda}$ as a function of  the scalar field mass $m^2/\Lambda$ (with $m<m_{cr}=0$) in the case of deSitter spacetime. Clearly $\Omega(m^2/\Lambda)<\vert m \vert$ as in the other cases where a cosmological horizon is present.}
\label{figomds} 
\end{figure}

In the pure deSitter case ($M=0$, $Q=0$), the potential $V_{*0}(r_*)$ is shown in Fig. \ref{splvrtds} for various values of $m^2/\Lambda$ and  may be obtained analytically as \cite{Du:2004jt}
\be
\begin{split}
V_{* 0}(r_*)&=\frac{m^2 -\frac{2}{3}\Lambda}{\cosh^2 \frac{r_*}{\sqrt{\frac{3}{\Lambda}}}}\simeq^{0<r_*\ll \sqrt{\frac{3}{\Lambda}}}\\
&\left( m^2-\frac{2}{3}\Lambda\right)+ \frac{\Lambda}{9}\left(2\Lambda-3m^2\right)r_*^2+\mathcal{O}(r_*^4)
\end{split}
\label{nlptwoeq35}
\ee
After a rescaling $r_* \sqrt{\Lambda} \rightarrow r_*$, $m^2/\Lambda \rightarrow m^2$ which practically amounts to setting $\Lambda=1$ it is obvious that the SSC is satisfied for $m^2>\frac{2}{3}$ which guarantees no instabilities for this range of $m^2$. Since there is only cosmological horizon in this case, the range of the tortoise coordinate is $r_*\in [0,+\infty]$. For $\Omega=0$ the Schrodinger-like equation to solve in this case takes the form

\be
\frac{du_0^2}{dr_*^2}-\frac{1}{\Lambda}V_*(r_*)u_0(r_*)=0
\label{nlptwoeq36}
\ee
Since the potential vanishes at $+\infty$ due to the cosmological horizon, the physically interesting (finite) boundary condition at $r_* \longrightarrow  +\infty$ is 

\be
u_0(r_*\rightarrow +\infty)=C 
\label{nlptwoeq37}
\ee
\be
u_0'(r_*\rightarrow +\infty)=0
\label{nlptwoeq38}
\ee
At the other boundary $r_*\rightarrow 0$ we have

\be
\frac{dr_*}{dr}=1\Longrightarrow r_*=r
\label{nlptwoeq39}
\ee
and due to Eqs. (\ref{nlptwoeq6}) and (\ref{nlptwoeq20})  for a finite scalar field at $r=0$  we must have

\be
\Psi_0(r\rightarrow 0)=0\Longrightarrow u_0(r\rightarrow 0)= u_0(r_*\rightarrow 0)=0
\label{nlptwoeq41}
\ee
Thus using Eqs. (\ref{nlptwoeq35}) and (\ref{nlptwoeq36})  it is straightforward to show that 
\be
 u_0(r_*\rightarrow 0)=r_*
\label{nlptwoeq44}
\ee
where we have used the normalization freedom to set the slope of the linear function to unity. Thus in this case, the physical boundary conditions are
\ba
u_0'(r_*\rightarrow 0)&=&1
\label{nlptwoeq45} \\
u_0 (r_*\rightarrow 0)&=&0
\label{nlptwoeq46}\\
u_0(r_*\rightarrow +\infty)&=&C
\label{nlptwoeq47}\\
u_0'(r_*\rightarrow +\infty)&=&0
\label{nlptwoeq48}
\ea
Solving  Eq. (\ref{nlptwoeq36})  corresponding to $\Omega=0$ from $r_*=0$ with the boundary conditions (\ref{nlptwoeq45}) and (\ref{nlptwoeq46}), we obtain (\ref{nlptwoeq47}) and (\ref{nlptwoeq48})  only for $m_{cr}=0$. Thus, despite of the negative effective Regge-Wheeler potential in the deSitter background, the tachyonic instabilities develop for the same range of $m^2$ as in the Minkowski space ($m^2<0$). It is straightforward to find the ground state eigenvalue and show that $\Omega(m^2/\Lambda)<\vert m \vert$ as in the case of other specetimes where a cosmological horizon is present (see Fig. \ref{figomds}).

\subsection{Pure Schwarzschild background}

\begin{figure*}
\begin{centering}
\includegraphics[width=0.97\textwidth]{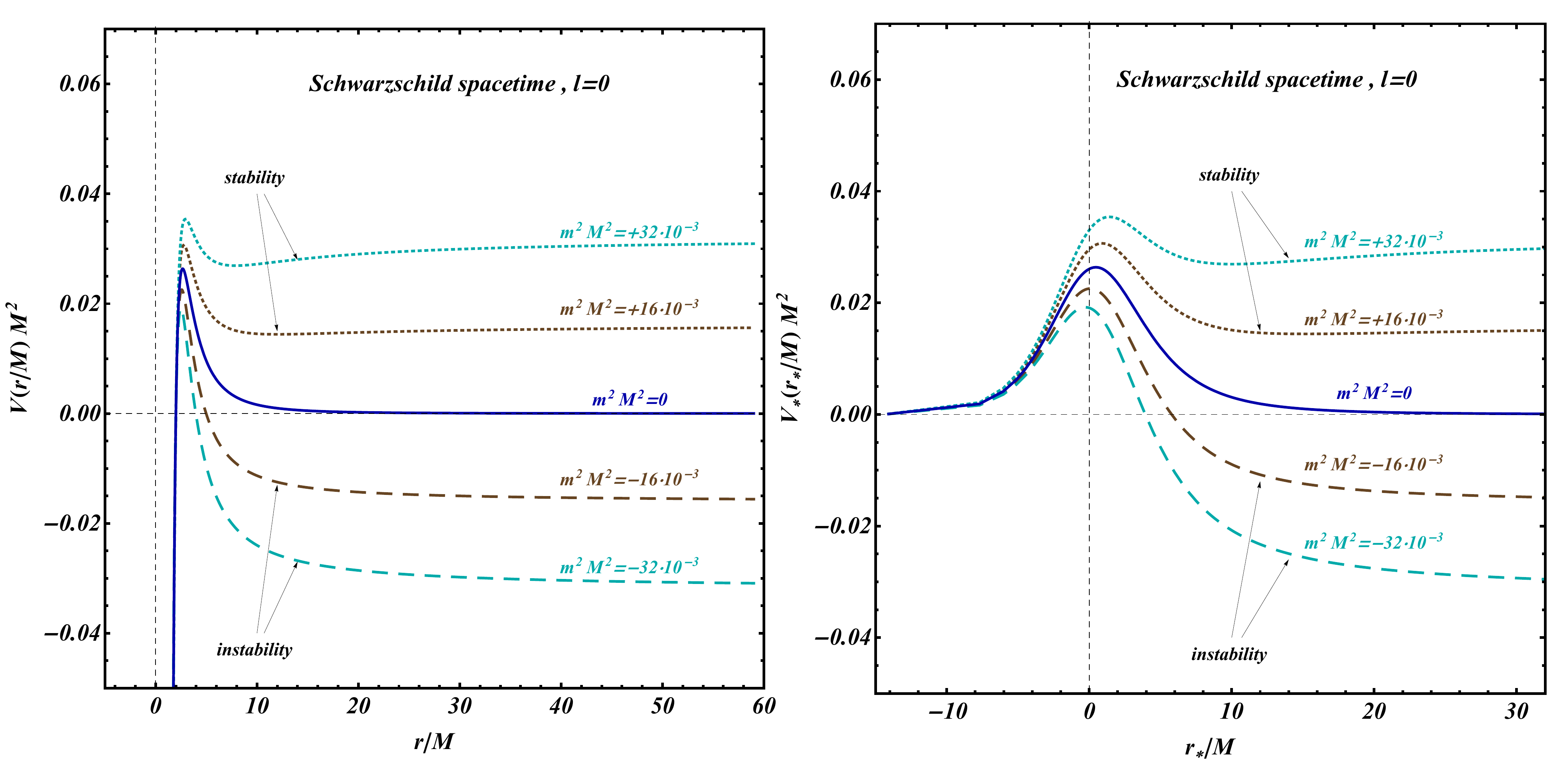}
\par\end{centering}
\caption{The  $m^2M^2$ dependent Regge-Wheeler dimensionless potentials  $V M^2$ (left panel) and  $V_* M^2$  (right panel) as a function of  $r/M$ and  $r_*/M$ respectively in the case of the Schwarzschild spacetime ($\Lambda=0$,  $\xi=0$) for angular scale $l=0$ . The  blue solid curves correspond to the critical value of the scalar field mass $m_{cr}^2M^2=0$. The dotted ($m^2M^2>0$) and dashed ($m^2M^2<0$) curves correspond to non-existence of bound states (stabilities) and existence of bound states (instabilities) respectively.} 
\label{splvrvrtsh} 
\end{figure*}

In the pure Schwarzschild background ($\Lambda=0$) we have \cite{Sibandze:2016agp,Li:2016sjq}
\ba
f(r)&=&1-\frac{2M}{r}
\label{nlptwoeq49a}\\
V(r)&=&\left(1-\frac{2M}{r}\right)\left(\frac{l(l+1)}{r^2}+\frac{2M}{r^3}+m^2\right) \label{nlptwoeq49b}\\
r_*(r)&=&r+2M\ln\left(\frac{r}{2M}-1\right)
\label{nlptwoeq49}
\ea
It is easy to see that in both the tortoise and the Schwarzschild coordinates the Regge-Wheeler potential $V_{*0}$ does not vanish asymptotically at $+\infty$. Instead we have (see also Fig. \ref{splvrvrtsh})
\be
\lim_{r_*\to +\infty}V_{*0}=m^2
\label{limvstar}
\ee
This implies that for $m^2<0$ the SIC implies instability since
\be 
\begin{split}
&\int_{-\infty}^{\infty} V(r_*)dr_*=\int_{r_H}^{\infty} V(r)dr=\\
&=\int_{r_H}^{\infty}\left(1-\frac{2M}{r}\right)\left(\frac{l(l+1)}{r^2}+\frac{2M}{r^3}+m^2\right)dr=-\infty<0
\end{split}
\label{nlptwoeq51}
\ee
Therefore for $m^2<0$ we have tachyonic instability just as in the Minkowski space. Similarly for $m^2>0$ we have $V(r)>0$ and $V_{*0}(r_*)>0$ which is the SSC (see also Fig. \ref{splvrvrtsh}) which secures that we have stability. Thus in the Schwarzchild backround, tachyonic instabilities develop for the same mass parameter range as for the Minkowski background.

In this case, for $m^2<0$, the boundary conditions (\ref{nlptwoeq25})-(\ref{nlptwoeq26}) become
\ba
u_0(r_*\rightarrow +\infty)&=& B e^{-i\sqrt{\vert m \vert^2-\Omega^2} r_*}
\label{nlptwoeq25n}\\
u_0(r_*\rightarrow -\infty)&=&A e^{\Omega r_*}
\label{nlptwoeq26n}
\ea
i.e. there are propagating waves towards $+\infty$ even for $m^2<0$. There are non-zero solutions satisfying these boundary conditions only for $\Omega\leq \vert m \vert$. This implies that the maximum growth rate of tachyonic istabilities in this case is the same as in flat space $\Omega=\vert m \vert$. This is due to the absence of a cosmological horizon.


\section{Conclusion-Discussion-Outlook} 
\label{conclusion}
We have shown that  tachyonic scalar instabilities of the KG equation have a  slower growth rate in  RN-dS/ SdS metric background compared to flat Minkowski space for all values of metric parameters where a cosmological horizons exists.  We have also identified the critical value of scalar field mass $m_{cr}^2$ that for $m^2<m_{cr}^2$ tachyonic instabilities develop and confirmed that $m_{cr}=0$ as in flat Minkowski spacetime.

The crucial property of the SdS spacetime that allows for this delayed growth of instabilities appears to be the presence of a cosmological horizon that forces the effective Regge-Wheeler potential to vanish at $+ \infty$ in tortoise coordinates even for negative scalar field mass $m^2$. Thus the $r_*$ range where the Regge-Wheeler potential is negative is limited  favoring increased eigenvalues and lower growth rate of instabilities.

This stabilizing effect of multiple horizons on tachyonic instabilities may have various interesting implications which include the following
\begin{itemize} 
    \item Tachyonic instabilities of $f(R)$ and scalar-tensor theories can get significantly delayed in backgrounds involving  cosmological horizons with possible implications for the development of preheating after inflation \cite{Allahverdi:2010xz,Felder:2000hj,He:2020ivk,Amin:2014eta}.
    \item Symmetry breaking phase transitions in field theory is based on the existence of tachyonic instabilities in a scalar field potential which lead the system towards a new vacuum state with less symmetry. In the context of a RN-dS background the delay of such tachyonic instabilities could have interesting effects in the evolution of phase transitions in the Early Universe with possible interesting observable effects related e.g. to the efficiency of the formation of topological defects \cite{Ye:1990na,Achucarro:1998ux}.
    \item The backreaction effects of the tachyonic instabilities on the gravitational background may lead to superradiance and scalarization effects \cite{Brito:2015oca,Winstanley:2001nx} in RN-dS spacetime in the same way that scattering processes lead to similar effects in these spacetimes.
    \item The consideration of scalar field potentials supporting topological or semilocal defects (e.g. electroweak strings \cite{James:1992zp}) may lead to interesting new stabilization mechanisms induced by a multihorizon gravitational background.
\end{itemize}

These implications open up a wide range of extensions of the present analysis. For example interesting extensions include the following:
\begin{itemize}
    \item Consideration of more general background metrics to investigate the existence and growth rate of tachyonic instability modes. Such backgrounds may include Kerr-Newman-deSitter spacetime \cite{Carter:1968rr,Stuchlik:1997gk,Winstanley:2001nx,Podolsky:2006px,Stuchlik:2008xk,Kraniotis:2014paa,Kraniotis:2016maw,Kraniotis:2018zmh,Kraniotis:2019ked} or corresponding higher dimension spacetimes, Gödel-like spacetime \cite{Konoplya:2011ag,Konoplya:2011hf} etc. 
    \item Investigate the effects of such delay of instabilities in the Early Universe and in particular during inflation and cosmological phase transitions \cite{Kibble:1976sj,Zurek:1985qw,Rajantie:2003xh,Kibble:1980mv} in the context of more general scalar field potentials beyond the KG equation.
    \item Investigate different types of perturbations (Dirac and gravitational) in multihorizon backgrounds and in the presence of tachyonic modes.
    \item Consider different types of boundary conditions corresponding to scattering processes (propagating waves at infinity) leading to evaluation of QNMs and scattering amplitudes (superradiance).
    \item Investigate the stability of semilocal and electroweak strings in strongly curved backgrounds including multihorizon metrics.
\end{itemize}

In conclusion, the interesting non-trivial effects of the gravitational background on the tachyonic scalar instabilities pointed out in the present analysis open up a wide range of new directions in the understanding of the dynamics of scalar fields in curved spacetimes.\\

\textbf{Supplemental Material:} The Mathematica file used for the numerical analysis and for construction of the figures can be found in \cite{suppl}.\\
 
\section*{ACKNOWLEDGEMENTS}
 This research is co-financed by Greece and the European Union (European Social Fund - ESF) through the Operational Programme ”Human Resources Development, Education and Lifelong Learning 2014-2020” in the context of the project ”Scalar fields in Curved Spacetimes: Soliton Solutions, Observational Results and Gravitational Waves” (MIS 5047648). This  article has also benefited from COST Action CA15117 (CANTATA), supported by COST (European Cooperation in Science and Technology).\\

\appendix
\section{Analytical form of SSC curves}
\label{Appendix}

The  above mentioned  Sufficient for Stability Criterion (SSC) is that the minimum of the Schrodinger potential should be larger than 0 (see Eq. (\ref{nlptwoeq34})). Thus by demanding that the minimum of the Schrodinger potential 
\be
V_{0 min}(r_{min})=0
\label{ssccurve}
\ee
we can obtain the analytical form of SSC curves for various values of $Q$ (see Fig. \ref{fig4new}). 
The SSC curve for $Q=0$ as function of $\xi$ takes the following analytical form
\begin{widetext}
\be 
m^2(\xi)M^2=\frac{2\left(g(\xi)-1\right)}{9g(\xi)^3}
\ee
where

\be
 g(\xi)=\frac{1}{\left(\sqrt{\xi^4-\xi^3}-\xi^2\right)^{\frac{1}{3}}}+\frac{\left(\sqrt{\xi^4-\xi^3}-\xi^2\right)^{\frac{1}{3}}}{\xi}
\ee
\end{widetext}
\bibliography{bibliography}              
 
\end{document}